\documentclass[a4paper, 11pt]{article}
\usepackage[utf8]{inputenc}
\usepackage[margin=2cm]{geometry}
\usepackage{graphicx, float}
\usepackage{amsmath}
\usepackage{amssymb}
\usepackage{color}
\usepackage{mathrsfs}
\usepackage{amsfonts}
\usepackage{array}
\usepackage{multirow}
\usepackage{multicol}
\usepackage{indentfirst}
\usepackage{enumitem} 
\usepackage{enumerate}
\usepackage{geometry}
\usepackage{tasks}
\usepackage[table,xcdraw]{xcolor}
\usepackage{secdot}
\usepackage[table]{xcolor}
\usepackage[english]{babel}
\usepackage{adjustbox}
\usepackage{lipsum}
\usepackage{authblk}
\usepackage{ulem}

\date{}
\begin{document}
\begin{titlepage}

\hfill\parbox{5cm} { }

\vspace{25mm}
\begin{center}
{\Large\bf Dynamical analysis in regularized $4D$ Einstein-Gauss-Bonnet gravity with non-minimal coupling}\\
 \vskip 1. cm
Bilguun Bayarsaikhan$^\text{a,b,}$\footnote{e-mail : ph.bilguun@gmail.com}, Sunly Khimphun$^\text{c,}$\footnote{e-mail : khimphun.sunly@rupp.edu.kh}, Phearun Rithy$^\text{c,}$\footnote{e-mail : phearunrain0@gmail.com}, and Gansukh Tumurtushaa$^\text{d,}$\footnote{e-mail : gansukh@jejunu.ac.kr}
  {\it$^a\,$Institute for Theoretical Physics, ELTE Eotvos Lorand University, Pazmany Peter setany 1/A, ~H-1117 Budapest, Hungary}\\
{\it $^b\,$Institute of Physics \& Technology, Mongolian Academy of Sciences, Ulaanbaatar 13330, Mongolia}\\
{\it $^c\,$Graduate School of Science, Royal University of Phnom Penh, Phnom Penh, Cambodia 12150}\\
{\it $^d\,$Department of Science Education, Jeju National University, Jeju, 63243, Korea}
\end{center}
\thispagestyle{empty}
\begin{abstract}
We investigate the regularized four-dimensional Einstein-Gauss-Bonnet ($4D$ EGB) gravity with a non-minimal scalar coupling function, which is an extension of the regularized $4D$ EGB theory. By introducing non-minimal coupling to the Gauss-Bonnet term, we demonstrate the additional contribution to the dynamical equations which is otherwise absent in the dimensionally-regularized theory. Furthermore, we analyze the stability of the system by using the dynamical system approach based on fixed points. Then, we consider the time evolution to investigate the history of the universe and constraint with observational data to obtain the cosmological parameters of the model.
\end{abstract}
\vspace{1cm}

\vspace{2cm}


\end{titlepage}

\section{Introduction}
\label{sec:intro}
Einstein's theory of gravity predicts the accelerating expansion of the late-time universe, which has been confirmed by the cosmological observations from type Ia Supernovae (SNIa)~\cite{SupernovaSearchTeam:1998fmf, SupernovaCosmologyProject:1998vns}. The expansion of the universe can be explained by the simplest model of the cosmological constant that currently gives the best fit with the current observational data from Cosmic Microwave Background (CMB) radiation, the measurement of Type Ia Supernovae (SNIa), Baryon Acoustic Oscillation (BAO), and $H_0$ measurement~\cite{Planck:2015fie, Planck:2018vyg}. 
However, some issues remain for the cosmological constant or $\Lambda$CDM model as encountered from the observations~\cite{Planck:2015fie, Bullock:2017xww, Bull:2015stt} and a seriously theoretical inconsistency~\cite{Carroll:1991mt, Weinberg:2000yb}, and for a review, one may refer to~\cite{Sahni:1999gb,DelPopolo:2016emo}.
 In these regards, several modified theories of gravity have been proposed to refine or provide alternatives. Without exhaustion, we list down some of the theories which deviate from the standard cosmological model:
  scalar-tensor theory \cite{Brans:1961sx, Horndeski:1974wa, Charmousis:2011bf, Perrotta:1999am, Khoury:2003aq}, vector-tensor \cite{Jacobson:2007veq}, tensor-tensor theory \cite{Rosen:1973, Milgrom:2009gv, Isham:1973bu}, massive gravity \cite{deRham:2010gu, deRham:2010ik, deRham:2010kj}, higher-order derivative theory, in particular, the f(R)-gravity for cosmology \cite{Starobinsky:1980te, Hwang:2001pu,Faulkner:2006ub,DeFelice:2010aj}, and higher-dimensional and string-motivated theory \cite{Randall:1999ee, Randall:1999vf, Dvali:2000hr, Apostolopoulos:2008ru, Lovelock:1971yv, Mueller-Hoissen:1985www}. For a nice review, one is referred to \cite{Clifton:2011jh}. Also, there are many dark energy models under these frameworks such as
 Chevallier Polarski Linder (CPL) model~\cite{Linder:2002et}, Holographic Dark Energy model and its constraint~\cite{Li:2004rb, Huang:2004wt, Zhang:2005hs, Zhang:2007sh, Chang:2005ph}, Generalized holographic cosmology via AdS/CFT constraint with low redshift~\cite{Khimphun:2020nkh}, and Chaplygin gas model~\cite{Kamenshchik:2001cp, Bento:2002ps} to name a few. One can refer to~\cite{Xu:2016grp} for model comparison using observational data from Planck 2015.

 It is well-known that Lovelock's theorem is a natural generalization to general relativity, which satisfies the diffeomorphism and local Lorentz invariance in $D$-dimensional space-time and, in particular, leads to second-order field equations~\cite{Lovelock:1971yv, Lovelock:1972vz, Lanczos:1938sf}. A special case of this is the second order in Lanczos-Lovelock gravity Lagrangian density so-called Einstein-Gauss-Bonnet gravity. 
Consequently, one possible modification of General Relativity is the $D\geq 5$ Gauss-Bonnet (GB) gravity which satisfies the properties in Lovelock's theorem, including the ghost-free~\cite{Zumino:1985dp}, natural generalization with Einstein, and cosmological terms~\cite{Zwiebach:1985uq}. For its study of higher dimensions in cosmology, one can refer to~\cite{Lorenz-Petzold:1988qmn, Andrew:2007xa, Atmjeet:2013yta}.
However, the GB term in $4D$ is topologically invariant, so it does not modify Einstein's theory since it does not contribute to gravitational dynamics. Nevertheless, the GB term can contribute to the $4D$ dynamical equations if one introduces a coupling function of a scalar field to the GB term~\cite{Rizos:1993rt, Kanti:1998jd}. Due to the extra scalar degree of freedom, this consideration leads to many gravity studies, including cosmic acceleration in inflationary models of the early universe, and tested against the observational data~\cite{Amendola:2005cr, Guo:2010jr, Koh:2014bka, Koh:2016abf, Koh:2018qcy, Tumurtushaa:2018agq, Chew:2020lkj}.

Recently, the $4D$ EGB model proposed in~\cite{Glavan:2019inb} introduces a scaling coupling constant $\alpha\rightarrow \alpha/(D-4)$, where $D$ is the space-time dimensions, and considers the limit $D\rightarrow 4$. This model gives rise to nontrivial contributions to the gravitational dynamics due to the extra contribution in the equations of motion (EoM) from the GB term as the consequence of the scaling coupling constant without introducing extra degrees of freedom. Notice that in this model the space-time was assumed to be continuous~\cite{Tomozawa:2011gp}. 
To be less exhaustive, there is an interesting work in cosmology adopting this model with the observational constraints which can resolve the coincidence problem~\cite{Wang:2021kuw} and also indicate that the re-scaling coupling constant of the model still needs the help of the cosmological constant to explain the accelerated expansion. For a review of the model, one may refer to~\cite{Fernandes:2022zrq}.

Due to its richness and peculiar consequences, comments and arguments on this novel $4D$ EGB theory have been raised~\cite{Gurses:2020ofy, Gurses:2020rxb, Arrechea:2020gjw, Arrechea:2020evj, Shu:2020cjw, Ai:2020peo, Mahapatra:2020rds}, and  the diffeomorphism invariant regularization is also considered in~\cite{Hennigar:2020lsl, Fernandes:2020nbq, Kobayashi:2020wqy, Lu:2020iav} to point out that the theory advertised in \cite{Glavan:2019inb} is the subclass of Horndeski theory. However, the arguments from Horndeski theory still rely on a $2+1$ degrees of freedom where higher dimensions are compactified and where in \cite{Glavan:2019inb} the space-time is based on the D-dimensional direct product and $D\to 4$ limit, and further investigation into this regularized tensor-scalar theory leading to a strong-coupling scalar field is studied in \cite{Kobayashi:2020wqy, Ma:2020ufk} and without the strong coupling field in \cite{Bonifacio:2020vbk}. Another investigation into the regularized $4D$ EGB theory is that looking at diffeomorphism invariant property reveals the inconsistency in the regularized $4D$ EGB theory~\cite{Arrechea:2020evj, Ai:2020peo, Mahapatra:2020rds}.
Nevertheless, the consistent study of $D \to 4$ EGB theory can be achieved up to spatially diffeomorphism invariance by using Hamiltonian formalism so-called minimally modified gravity theory; otherwise, an extra degree of freedom is required~\cite{Aoki:2020lig, Aoki:2020iwm}. Therefore, we will proceed by introducing an additional degree of freedom but still exploiting the regularization scheme which is expected to render the finite term that contributes to the dynamics. That is, we will scale $\alpha\rightarrow \alpha/(D-4)$ and take the limit $D\to 4$ after introducing the non-minimal coupling function to the GB term.  
 In this regard, we are interested in generalizing this concept by combining the role of the non-minimal coupling function often used in ordinary $4D$ gravity with the scaling coupling constant advocated in the  regularized $4D$ EGB gravity together. This seems to be a redundant consideration at first because the role of scaling coupling is to render the term of the overall factor ($D-4$) coming from the GB term so that these higher curvature terms become relevant, which is also  the role of the non-minimal coupling function. As a result, in the current study, we accept the necessity of the extra scalar degree of freedom in the framework of regularized $4D$ EGB.   

In this works, we study the extension of regularized $4D$ EGB theory by introducing the non-minimal coupling to the GB term and redefining the coupling function $$\xi(\phi) \to \xi^{(D-4)}(\phi),$$ with flat FRW metric in D-dimensions then take the limit $D \to 4$. The motivation of the form of coupling function is obvious so that divergence of the terms associated to the scaling of $\alpha$ remains well-defined in the EoM. From this consideration, it is obvious that the action could be ambiguous but not  ill-defined since this happens to the original regularized $4D$ EGB where its action might be investigated further  \cite{Hennigar:2020lsl, Fernandes:2020nbq, Kobayashi:2020wqy, Lu:2020iav}; this issue is not our concern here. From this, we apply the dynamical system approach (DSA) \cite{Ivanov:2011vy, Boehmer:2014vea, Granda:2017oku, SantosDaCosta:2018bbw, Bahamonde:2017ize} with the exponential of the potential and the coupling function of the scalar field into the dynamical equations of the model. We find the fixed points and analyze the stability of the autonomous system equations including the consideration of the evolution of the phase space universe. Particularly we consider the evolution of the cosmological parameters as the function of the red-shift and constraint on the potential parameter with the observational bounds from CMB, BAO, SnIa, and $H_0$ measurement. 

This article is organized as follows. In section~\ref{sec:sec2}, we introduce our setup and derive the EoM for our model. As a result, there appear two types of additional contributions in the $4D$ EoM each coming from the non-minimal coupling of the scalar field to the GB term and the scaling coupling constant of the GB term. Thus, it is imperative for us to study their respective role and their dynamic evolution throughout the cosmic history of the universe. To investigate the dynamical evolution of the universe for our model, we rewrite the background EoM in the autonomous system form in section~\ref{sec:sec3} and apply the dynamical system approach, which gives a robust description of the cosmic history based on the existence of critical points and their stability. In section~\ref{sec:sec4}, we obtain not only the critical fixed points of the system but also consider the stability of the universe at each point. We solve the dynamical equations for a broader time scale to better understand how the aforementioned GB contributions evolve, especially at the late time, and provide our numerical result with their implications. A summary of our results in this paper and a conclusion are given in section~\ref{sec:sec5}.
\section{The setup and Equations of Motion}\label{sec:sec2}

In this work, we consider the action in $D$-dimensional space-time as 
 follow,
\begin{equation}
S =\int\mathrm{d}^D x\sqrt{-g}\left[\frac{1}{2\kappa^2}R-\frac{1}{2}g^{\mu\nu}\partial_\mu\phi\partial_\nu\phi-V(\phi)-\frac{\alpha}{2}\xi(\phi)\mathcal{G} \right]+S_{\text{m,r}}, \label{eq:GB+non-minimal}
\end{equation}
where $R$, $\alpha$, $V(\phi)$, $\xi(\phi)$ are Ricci scalar, a coupling constant, potential function and the coupling function of scalar field respectively. $$\mathcal{G} \equiv R^2-4R_{\mu\nu}R^{\mu\nu}+R_{\mu\nu\rho\sigma}R^{\mu\nu\rho\sigma}\,,$$ is GB term, $S_{\text{m,r}}$ represents the standard matter and radiation components, $\kappa^2=8\pi G$, and $G$ is Newtonian constant. From this action, after obtaining the EoM, we re-scale $\alpha\to\alpha/(D-4)$ and then take the limit $D\rightarrow 4$. The problem arises when we introduce $\xi(\phi)$ couples to GB term since the expected finite terms from $\xi(\phi)$ which appear in the EoM contain no factor of ($D-4$). As a result, $1/(D-4)$ from the scaling $\alpha$ will diverge. We propose the redefining the coupling function $$\xi(\phi)\rightarrow\xi^{(D-4)}(\phi),$$ first and then take the limit $D\rightarrow 4$. Therefore, denominator $1/(D-4)$ will be canceled out with the overall factor ($D-4$) coming from the redefined coupling function. From the least-action principle, the field equations after scaling $\alpha$ and redefining $\xi(\phi)$ are
\begin{align}\label{eq:field}
&R_{\mu\nu}-\frac{1}{2}g_{\mu\nu}R=\kappa^2\left(T^{(\phi)}_{\mu\nu}+T^{\text{(m,r)}}_{\mu\nu}+T^{\text{(GB)}}_{\mu\nu}\right), \\
&\Box \phi-V_\phi-\frac{\alpha}{2(D-4)}(D-4)\xi^{(D-5)}\xi_\phi \mathcal{G}=0,
\end{align}
where $\Box \equiv \nabla_\mu \nabla^\mu$,  $V_\phi=dV/d\phi$, $\xi_\phi=d\xi/d\phi$, and $T^{(\phi)}_{\mu\nu},~T^{\text{(m,r)}}_{\mu\nu},~T^{\text{(GB)}}_{\mu\nu}$ are stress-energy-momentum tensor of scalar field, GB term and matter term respectively, which we find
\begin{align}
T^{(\phi)}_{\mu\nu}&=\partial_\mu \phi \partial_\nu \phi-\frac{1}{2}g_{\mu\nu}\left(g^{\rho \sigma}\partial_\rho \phi \partial_\sigma \phi  +2V \right), \\
T^{(\text{m,r})}_{\mu\nu}&=(\rho+P)U_\mu U_\nu+ P g_{\mu\nu},
\end{align}
{\footnotesize\begin{align}
T^{(\text{GB})}_{\mu\nu}=&\frac{\alpha}{(D-4)}\Biggl\{\xi^{(D-4)}\left(-4R_\mu~ ^\alpha R_{\nu \alpha}+2R R_{\mu\nu}-4R^{\alpha\beta}R_{\mu\alpha\nu\beta} +2R_\mu~^{\alpha\beta\gamma} R_{\nu\alpha\beta\gamma} -\frac{1}{2} g_{\mu\nu} \mathcal{G} \right) \nonumber\\
&+(D-4)(D-5)\xi_\phi^2\xi^{(D-6)} \Biggl[ g_{\mu\nu}\left(2R  \nabla^\alpha \phi \nabla_\alpha \phi-4R_{\alpha\beta}\nabla^\alpha \phi \nabla^\beta\phi  \right)-4R_{\mu\nu}\nabla^\alpha \phi \nabla_\alpha \phi \nonumber \\
&+4R_{\mu\alpha\nu\beta}\nabla^\alpha \phi \nabla^\beta \phi +4R_{\nu\alpha}\nabla^\alpha \phi \nabla_\mu \phi +4R_{\mu\alpha} \nabla^\alpha \phi \nabla_\nu \phi-2R\nabla_\mu \phi \nabla_\nu \phi  \Biggr]+(D-4)\xi^{(D-5)}\nonumber\\
&\quad \times\Biggl[\xi_\phi \Bigl(-4R_{\mu\nu}\Box \phi+2R_{\mu\alpha\nu\beta} \nabla^\beta \nabla^\alpha \phi+2R_{\mu\beta\nu\alpha} \nabla^\beta \nabla^\alpha \phi+g_{\mu\nu}\Bigl(2R\Box \phi-4R_{\alpha\beta}\nabla^\beta \nabla^\alpha \phi   \Bigr) \nonumber\\
&+4R_{\nu\alpha}\nabla_\mu \nabla^\alpha \phi+4R_{\mu\alpha}\nabla_\nu \nabla^\alpha \phi-2R\nabla_\nu \nabla_\mu \phi \Bigr)+\xi_{\phi\phi}\Bigl(-4R_{\mu\nu}\nabla^\alpha \phi \nabla_\alpha \phi   +4R_{\mu\alpha\nu\beta}\nabla^\alpha \phi \nabla^\beta \phi \nonumber\\
&+g_{\mu\nu}\Bigl( 2R\nabla^\alpha \phi \nabla_\alpha \phi-4R_{\alpha\beta}\nabla^\alpha \nabla^\beta \phi  \Bigr)+4R_{\nu\alpha}\nabla^\alpha \phi \nabla_\mu \phi  +4R_{\mu\alpha}\nabla^\alpha \phi \nabla_\nu \phi -2R\nabla_\mu \phi \nabla_\nu \phi \Bigr)  \Biggr]  \Biggr\}, 
\end{align}} where $U_\mu=(1,0,0,0)$ is four-velocity and $\rho,~P$ are the energy density and pressure in perfect fluid respectively. The homogeneous and isotropic universe spatially flat $k=0$ of Friedmann-Robertson-Walker (FRW) metric in $D$-dimensions is given
\begin{equation}
\begin{split}
\text{d}s^2&=-\text{d}t^2+a^2(t)(\text{d}\chi_1^2+\text{d}\chi_2^2+\text{d}\chi_3^2+...+\text{d}\chi_{D-1}^2),
\end{split}
\end{equation}
where $a(t)$ is a scale factor. Notice that above line element is a $D$-dimensional product space. By solving the field equations with flat FRW metric in arbitrary dimensions we can write the EoM in general dimensions for the components $tt$, $ii$, and scalar field equation as the following
\begin{align}
&\frac{(D-2)(D-1)}{2\kappa^2}H^2 =\frac{1}{2}\dot{\phi}^2+V+\rho_\text{m}+\rho_\text{r}+2\alpha(D-3)(D-2)(D-1)\dot{\xi}\xi^{(D-5)}H^3\nonumber\\
&\qquad +\frac{1}{2}\alpha(D-3)(D-2)(D-1)\xi^{(D-4)}H^4, \label{eq:timecomponent}\\
&-\left[\frac{(D-3)(D-2)}{2\kappa^{2}}H^2+\frac{(D-2)}{\kappa^{2}}\frac{\ddot{a}}{a} \right] = \frac{1}{2}\dot{\phi^2}-V+p_\text{r}-4\alpha(D-3)(D-2)\dot{\xi}\xi^{(D-5)}H\frac{\ddot{a}}{a}\nonumber\\
&\qquad -2\alpha(D-3)(D-2)\xi^{(D-4)}H^2\frac{\ddot{a}}{a}-2\alpha(D-3)(D-2)H^2\left[(D-5)\dot{\xi}^2\xi^{(D-6)}+\ddot{\xi}\xi^{(D-5)} \right]\nonumber\\
&\qquad \qquad -2\alpha(D-4)(D-3)(D-2)\dot{\xi}\xi^{(D-5)}H^3-\frac{1}{2}\alpha(D-3)(D-2)(D-5)\xi^{(D-4)}H^4, \\
&0=\ddot{\phi}+V_\phi+(D-1)\dot{\phi}H+2\alpha(D-3)(D-2)(D-1)\xi_\phi\xi^{(D-5)}H^2\frac{\ddot{a}}{a}\nonumber \\
&\qquad +\frac{\alpha}{2}(D-4)(D-3)(D-2)(D-1)\xi_\phi\xi^{(D-5)}H^4,\label{eq:scalarfieldcomponent}
\end{align}
where $V_\phi=dV/d\phi$, $H \equiv \dot{a}/a$, denote the derivative of potential with respect to scalar field and Hubble's parameter respectively. A dot mean derivative with respect to the cosmic time,  $\rho_\text{m}$ and $\rho_\text{r}$ are the energy density of matter corresponding to pressure  $p_\text{m}=0$ and radiation corresponding to $p_\text{r}=\rho_\text{r}/3$ respectively. Then, determining the limit $D\to4$ for the Eqs. (\ref{eq:timecomponent})-(\ref{eq:scalarfieldcomponent}), we derive the explicit dynamical equations in $4$-dimensions as
\begin{align}
& 3H^2 =\kappa^2\left(\rho_\text{m}+\rho_\text{r}+\frac{1}{2}\dot{\phi}^2+V+12\alpha\frac{\dot{\xi}}{\xi}H^3+3\alpha H^4  \right), \label{eq:Novel$4D$FirstFried}\\
& 2\dot{H}+3H^2 =-\kappa^2\left[\frac{\rho_\text{r}}{3}+\frac{\dot{\phi}^2}{2}-V-8\alpha\frac{\dot{\xi}}{\xi} H(\dot{H}+H^2)-3\alpha H^4-4\alpha H^2\dot{H}-4\alpha\left(-\frac{\dot{\xi}^2}{\xi^2}+\frac{\ddot{\xi}}{\xi} \right)H^2 \right],\label{eq:Novel$4D$SecFried}\\
&\ddot{\phi}+V_\phi+3\dot{\phi}H+12\alpha\frac{\xi_\phi}{\xi}H^2\left(\dot{H}+H^2\right)=0,\label{eq:ScalarField}
\end{align}
where $\xi_\phi=d\xi/d\phi$, $\xi_{\phi\phi}=d^2\xi/d\phi^2$ and $\dot{\xi}=\xi_\phi \dot{\phi}$ and in (\ref{eq:Novel$4D$FirstFried}), the quartic term in Hubble parameter is the term that appears in \cite{Glavan:2019inb}, but our equations add the cubic terms in Hubble parameter and terms associated with $\xi(\phi)$ due to the regularization of the non-minimal coupling function to the GB term. These additional terms might give explanations of the history of the universe from the early to the late-time universe. The continuity equation of radiation density is
\begin{align}
\dot{\rho_\text{r}}+4H\rho_\text{r}=0.
\end{align}
By using the redefinition of some quantities, we can rewrite Eqs. (\ref{eq:Novel$4D$FirstFried}) and (\ref{eq:Novel$4D$SecFried}) as
\begin{align}
3H^2&=\kappa^2\left(\rho_\text{m}+\rho_\text{r}+\rho_\phi \right),\\
2\dot{H}+3H^2&=-\kappa^2\left(\frac{1}{3}\rho_\text{r}+p_\phi  \right),
\end{align}
where the energy density and the pressure of the scalar field is defined as
\begin{align}
\rho_\phi&=\frac{1}{2}\dot{\phi}^2+V+12\alpha\frac{\dot{\xi}}{\xi}H^3+3\alpha H^4, \label{eq.rho}\\
p_\phi&=\frac{\dot{\phi}^2}{2}-V-8\alpha\frac{\dot{\xi}}{\xi} H(\dot{H}+H^2)-3\alpha H^4-4\alpha H^2\dot{H}-4\alpha\left(-\frac{\dot{\xi}^2}{\xi^2}+\frac{\ddot{\xi}}{\xi} \right)H^2. \label{eq.pressure}
\end{align}
The components of $\rho_\phi$ and $p_\phi$ satisfy the conservation equation and the Eq. (\ref{eq:ScalarField}) can also be rewritten as
\begin{align}
\dot{\rho_\phi}+3H(1+w_\phi)\rho_\phi=0,
\end{align}
where the equation of state (EoS) for the scalar field can be defined as $w_\phi\equiv p_\phi/\rho_\phi$, and the effective EoS can also be defined as
\begin{align}
w_\text{eff}\equiv -1-\frac{2\dot{H}}{3H^2}. \label{eq.state}
\end{align}
\section{Dynamical System Approach} \label{sec:sec3}
To understand the cosmological dynamics of this system, we apply the dynamical system approach studied in \cite{Ivanov:2011vy, Boehmer:2014vea, Granda:2017oku, SantosDaCosta:2018bbw, Bahamonde:2017ize}. First, we define the dimensionless variables in order to conveniently investigate the dimensionless dynamical equations. Then, we find the fixed points of the autonomous system equations and analyze the stability. From that, we work further on the time evolution of cosmological parameters. We can rewrite (\ref{eq:Novel$4D$FirstFried}) as
\begin{align}
1&=\frac{\kappa^2 \rho_\text{m}}{3H^2}+\frac{\kappa^2 \rho_\text{r}}{3H^2}+\frac{\kappa^2\dot{\phi}^2}{6H^2}+\frac{\kappa^2 V}{3H^2}+\frac{ 4\alpha \kappa^2\dot{\xi}H}{\xi}+\alpha \kappa^2 H^2,
\end{align}
and define the dimensionless variables as
\begin{align}
x&=\frac{\kappa \dot{\phi}}{\sqrt{6}H}, ~~y=\frac{\kappa \sqrt{V}}{\sqrt{3}H},~~\alpha_{\text{GB}}=\sqrt{\alpha}\kappa H ,~~\mu=-\frac{\sqrt{6}\xi_\phi}{\kappa \xi},~~\lambda=-\frac{V_\phi}{\sqrt{6} \kappa V},~~\epsilon=-\frac{\dot{H}}{H^2}.\label{eq:variables}
\end{align}
From (\ref{eq:variables}), the Friedmann equation (\ref{eq:Novel$4D$FirstFried}) can be written as
\begin{align}
1&=\Omega_\text{m}+\Omega_\text{r}+\Omega_\phi,\\
\Omega_\text{m}&= \frac{\kappa^2 \rho_\text{m}}{3H^2},\\
\Omega_\text{r}&= \frac{\kappa^2 \rho_\text{r}}{3H^2},\\
\Omega_\phi&=x^2+y^2+\alpha_{\text{GB}}^2-4x \alpha_{\text{GB}}^2 \mu. \label{eq:wholedimensionless}
\end{align}
We also introduce the variable $N=\text{ln}a$ and $dN=Hdt$ so that by taking the derivative of these with respect to $N$, we derive the following system of equations
\begin{align}
\frac{dx}{dN}&=\frac{\kappa\ddot{\phi}}{\sqrt{6}H^2}-\frac{\kappa\dot{\phi}\dot{H}}{\sqrt{6}H^3}=x(\epsilon-\delta),\\
\frac{dy}{dN}&=\frac{\kappa V_\phi \dot{\phi}}{2\sqrt{3}\sqrt{V}H^2}-\frac{\kappa \sqrt{V}\dot{H}}{\sqrt{3}H^3}=y (\epsilon-3x\lambda),\\
\frac{d\alpha_{\text{GB}}}{dN}&=\frac{\sqrt{\alpha} \kappa \dot{H}}{H}=-\alpha_{\text{GB}}\epsilon,\\
\frac{d\Omega_\text{r}}{dN}&=-\frac{2\kappa^2 \rho_\text{r}\dot{H}}{3H^4}+\frac{\kappa^2 \dot{\rho}_\text{r}}{3H^3}=-4\Omega_\text{r}+2\Omega_\text{r}\epsilon,\\
\frac{d\mu}{dN}&=\frac{\sqrt{6}\xi_\phi^2 \dot{\phi}}{\kappa \xi^2 H}-\frac{\sqrt{6}\xi_{\phi\phi}\dot{\phi}}{\kappa \xi H}=x\mu^2(1-\Delta),\\
\frac{d\lambda}{dN}&=\frac{V_\phi^2 \dot{\phi}}{\sqrt{6}\kappa V^2 H}-\frac{V_{\phi\phi}\dot{\phi}}{\sqrt{6}\kappa V H}=6x \lambda^2(1-\Gamma),
\end{align}
where 
\begin{align}\label{eq:GammaDelta}
\Gamma\equiv\frac{V_{\phi\phi}V}{V_\phi^2}, \quad \Delta\equiv\frac{\xi_{\phi\phi}\xi}{\xi_\phi^2}\,,
\quad
\delta\equiv-\frac{\ddot{\phi}}{\dot{\phi}H}=3-\frac{3y^2 \lambda}{x}-\frac{2\alpha_{\text{GB}}^2 \mu}{x}+\frac{2\alpha_{\text{GB}}^2\mu \epsilon}{x}.
\end{align}
The autonomous system equations can also be written as
\begin{align}
\frac{dx}{dN}&=-3x+3y^2\lambda+2\alpha_{\text{GB}}^2 \mu-2\alpha_{\text{GB}}^2\mu \epsilon+x\epsilon, \label{eqx}\\
\frac{dy}{dN}&=-3x y \lambda+y \epsilon, \label{eqy}\\
\frac{d\alpha_{\text{GB}}}{dN}&=-\alpha_{\text{GB}}\epsilon,\label{eqz}\\
\frac{d\Omega_\text{r}}{dN}&=-4\Omega_\text{r}+2\Omega_\text{r}\epsilon, \label{eqomega}\\
\frac{d\mu}{dN}&=x\mu^2(1-\Delta),\label{eqmu}\\
\frac{d\lambda}{dN}&=6x \lambda^2(1-\Gamma)\,,\label{eqlambda}
\end{align}
where it is worth noting that these equations are invariant under $y\to -y$. 
From second Friedmann's equation (\ref{eq:Novel$4D$SecFried}), we derive 
\begin{align}\label{eq:epsilon}
\epsilon\equiv-\frac{\dot{H}}{H^2}
=&{}\frac{1}{2+4\alpha_{\text{GB}}^2(-1+2x\mu+2\alpha_{\text{GB}}^2 \mu^2)}\Bigl[3-3\alpha_{\text{GB}}^2-4x \alpha_{\text{GB}}^2 \mu+8\alpha_{\text{GB}}^4 \mu^2\nonumber\\
&{}+3y^2(-1+4\lambda\alpha_{\text{GB}}^2 \mu)+3x^2+4x^2 \alpha_{\text{GB}}^2 \mu^2 (1-\Delta)+\Omega_\text{r}  \Bigr].
\end{align}
Eq.~(\ref{eq:epsilon}) diverges for
\begin{align}\label{eq:divergence}
    x\equiv x_\text{div}=\frac{-1+2\alpha_\text{GB}^2-4\alpha_\text{GB}^4\mu^2}{4\alpha_\text{GB}^2\mu}\,,
\end{align}
which is unphysical. To avoid from this unphysical divergence the denominator must not be zero, \emph{i.e.,} $x\neq x_\text{div}$.   
From Eqs. (\ref{eq.rho}), (\ref{eq.pressure}) and (\ref{eq.state}), we derive the EoS parameters for both the scalar field and the effective as 
{\small
\begin{align}
w_\phi=&{}\frac{1}{3(x^2+y^2+\alpha_{\text{GB}}^2-4x \alpha_{\text{GB}}^2\mu)(1+4\alpha_{\text{GB}}^4\mu^2-2\alpha_{\text{GB}}^2+4x \alpha_{\text{GB}}^2\mu)}\left[\left(-3y^2+12y^2 \alpha_{\text{GB}}^2 \lambda \mu \right)\right.\nonumber\\
&{}+x^2\left.\left(3+4 \alpha_{\text{GB}}^2 \mu^2(1-\Delta)\Bigr) -4x \alpha_{\text{GB}}^2 \mu(4+\Omega_\text{r})+\alpha_{\text{GB}}^2\Bigl(3+2\Omega_\text{r}-4\alpha_{\text{GB}}^2 \mu^2(1+\Omega_\text{r})  \right)   \right]\,,\\
w_\text{eff} &\equiv -1-\frac{2\dot{H}}{3H^2}=-1+\frac{2\epsilon}{3}.
\end{align}}
\section{The Stability of Fixed Points and Numerical Results}\label{sec:sec4}
In this section, we investigate the stability of the autonomous system described by Eqs.~(\ref{eqx})--(\ref{eqlambda}) for the potential $V(\phi)$ and the coupling function $\xi(\phi)$ are given as
\begin{align}\label{eq:potandcfun}
V(\phi)=V_0 e^{-\sqrt{6}\kappa \lambda \phi},~~~\text{and}~~~\xi(\phi)=\xi_0 e^{-\kappa \mu \phi/\sqrt{6}}.
\end{align} 
where $\lambda$ and $\mu$ are arbitrary constants. It is straightforward to show that $\Delta=1$ and $\Gamma=1$ by substituting Eq.~(\ref{eq:potandcfun}) into Eq.~(\ref{eq:GammaDelta}). Thus, the autonomous system equations Eqs.~(\ref{eqx})--(\ref{eqomega}) are sufficient for further study. We first obtain fixed points and, then, check the stability of the system at those fixed points through the linear stability theory, which is based on analyzing the eigenvalues of the Jacobian matrix for the autonomous system equations. In particular, a sign of the real part of the eigenvalues of the Jacobian matrix plays an important role in determining the system's stability. If the real parts of all the eigenvalues are negative, it implies that the fixed point is stable or attractors. In contrast, the real parts of all the eigenvalues with a positive sign indicating that the fixed point is unstable or repellers. The fixed point is identified as a saddle point if some of the eigenvalues are negative while others are positive. 

\begin{figure}[t!]
    \centering
    \includegraphics[width=0.48\textwidth]{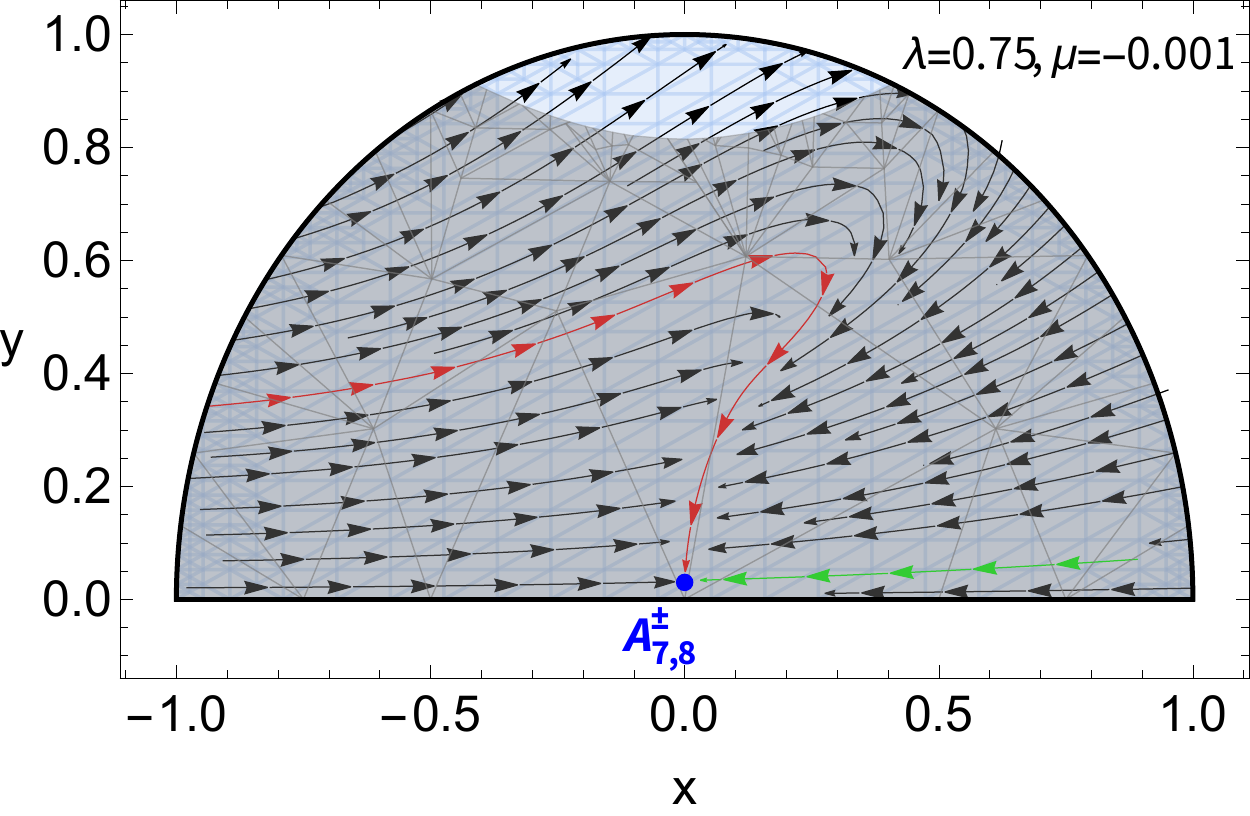}\quad 
    \includegraphics[width=0.48\textwidth]{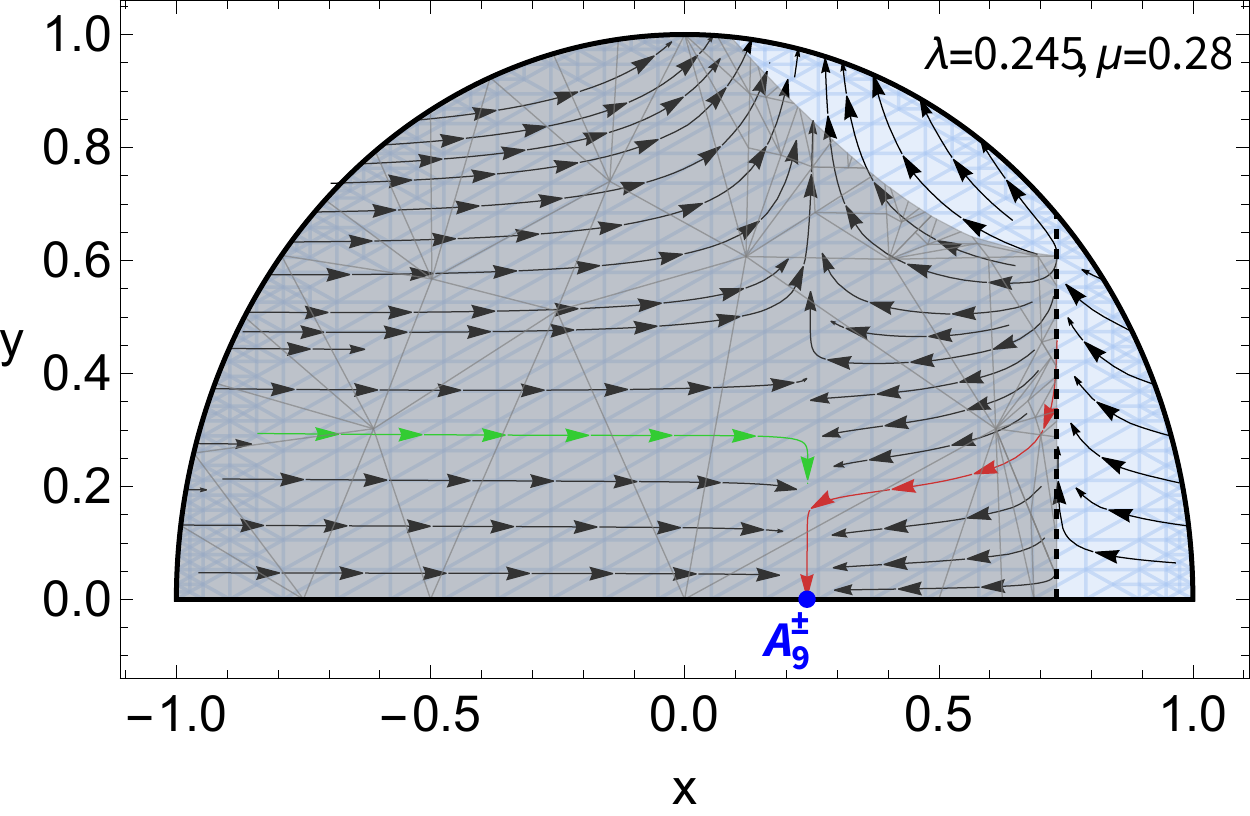}\\
    \includegraphics[width=0.48\textwidth]{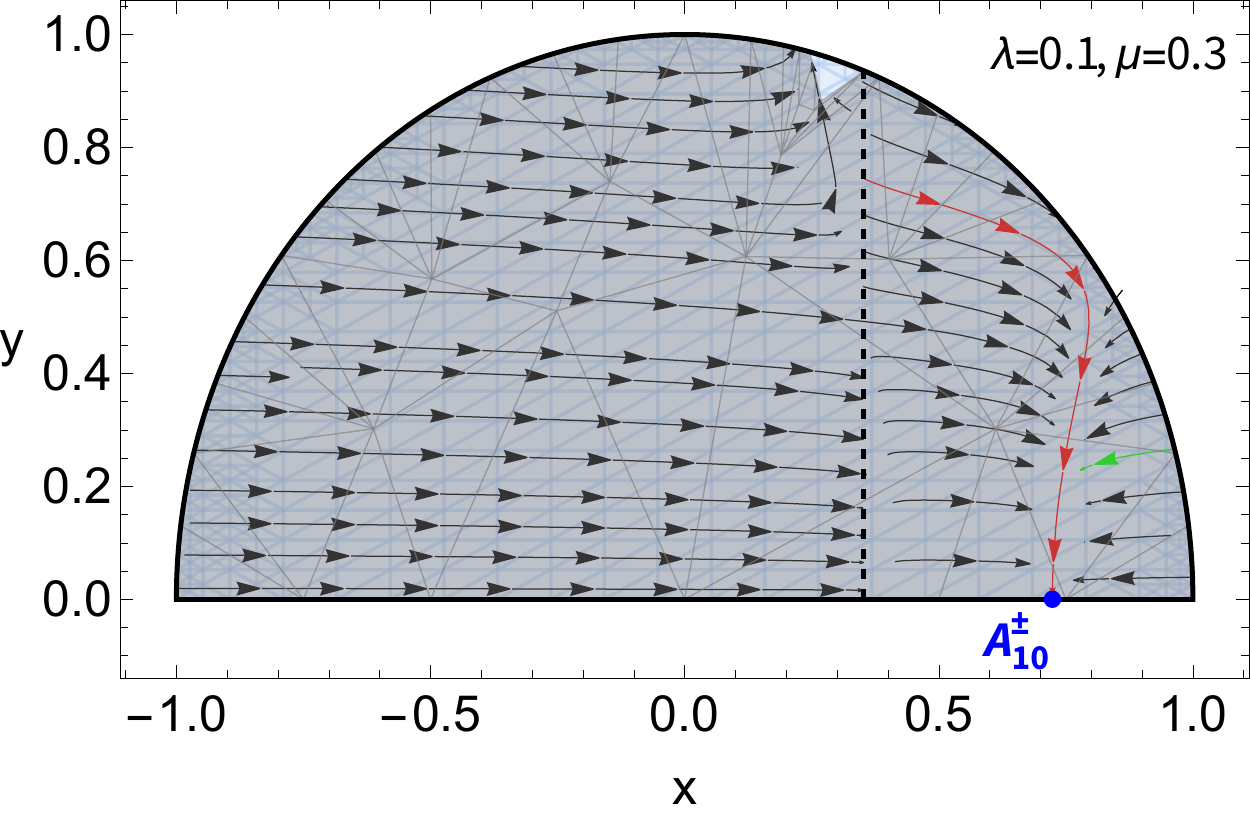}
    \caption{Phase space portraits of the autonomous system on a $x$ vs. $y$ plane. The acceleration of the universe is possible within the gray-shaded region. The green and red lines are particular attractor trajectories toward the fixed points $A_{7,8}^\pm$, $A_{9}^\pm$, and $A_{10}^\pm$ when their stability conditions are satisfied. The dashed lines at $x=x_\text{div}$ indicate the divergence of dynamical equations, see Eq.~(\ref{eq:divergence}). }
    \label{fig:fig1}
\end{figure}

We list all fixed points of our system in Table~\ref{table:nonlin1} and present their eigenvalues, as well as stability analyses of each fixed point, in Table~\ref{table:nonlin2}. In Table~\ref{table:nonlin1}, all the fixed points are arranged into two groups: (i.) the GR fixed points, including from $A^{\pm}_{1}$ to $A^{\pm}_{6}$, where $\alpha_\text{GB}=0$ and (ii.) the GB fixed points, from $A^{\pm}_{7}$ to $A^{\pm}_{10}$, where $\alpha_\text{GB}\neq0$. Since the stability of the GR fixed points is well investigated in the literature~\cite{Boehmer:2014vea, Bahamonde:2017ize}, we focus on understanding the stability of the GB fixed points in this work. In Figure~\ref{fig:fig1}, we plot the phase portraits of the autonomous system in the ``$x$--$y$'' plane. We only present the positive ``$y$''-axis because the background equations are symmetric under the change of $y\rightarrow-y$. These symmetric properties of the EoM imply that the $A_7^\pm$ and $A_8^\pm$ fixed points in  Table~\ref{table:nonlin1} are basically the same points.

\begin{table}[t!]
\centering 
\vspace{3mm}
\begin{adjustbox}{width=\textwidth}
\begin{tabular}{c| c c c | c c c| c| c}
\hline\hline  
Points & $x$ & $y$ & $\alpha_{\text{GB}}$ & $\Omega_r$&$\Omega_m$&$\Omega_\phi$& Existence &  $\omega_\text{eff}$=$\omega_\phi$ \\ 
\hline 
$A_1^\pm$ & $\frac{1}{2\lambda}$ & $\pm\frac{1}{2\lambda}$ & 0 & 0 & $1-\frac{1}{2\lambda^2}$ & $\frac{1}{2\lambda^2}$ & $\lambda^2>1/2$ & 0 \\
$A_2^\pm$ & $\frac{2}{3\lambda}$ & $\pm\frac{\sqrt{2}}{3\lambda}$ & 0 & $1-\frac{2}{3\lambda^2}$ & 0 & $\frac{2}{3\lambda^2}$ & $\lambda^2>2/3$ & $1/3$\\
$A_3^\pm$ & $\pm1$ & 0 & 0 & 0 & 0 & 1 & $\forall\lambda$ & 1 \\
$A_4$ & 0 & 0 & 0 & 0 & 1 & 0 & $\forall\lambda$ & 0 \\
$A_5$ & 0 & 0 & 0 & 1 & 0 & 0 & $\forall\lambda$ & $1/3$ \\
$A_6^\pm$ & $\lambda$ & $\pm\sqrt{1-\lambda^2}$ & 0 & 0 & 0 & 1 & $\lambda^2<1$ & $-1+2\lambda^2$\\ \hline
$A_7^\pm$ & 0 & $-\sqrt{\frac{2\mu}{2\mu-3\lambda}}$ & $\pm\sqrt{\frac{3\lambda}{3\lambda-2\mu}}$ & 0 & 0 & 1 & $(\mu\geq0, \lambda<0)$ or $(\mu\leq0, \lambda>0)$ & $-1$ \\
$A_8^\pm$ & 0 & $\sqrt{\frac{2\mu}{2\mu-3\lambda}}$ & $\pm\sqrt{\frac{3\lambda}{3\lambda-2\mu}}$ & 0 & 0 & 1 & $(\mu\geq0, \lambda<0)$ or $(\mu\leq0, \lambda>0)$ & $-1$ \\
$A_9^\pm$ & $\frac{3-\sqrt{9-80\mu^2}}{20\mu}$ & 0 & $ \pm\sqrt{\frac{6}{3+\sqrt{9-80\mu^2}}}$ & 0 & 0 & 1 & $-\frac{3}{4\sqrt{5}}\leq\mu\leq\frac{3}{4\sqrt{5}}, \mu\neq0$ & $-1$ \\
$A_{10}^\pm$ & $\frac{3+\sqrt{9-80\mu^2}}{20\mu}$ & 0 & $\pm\sqrt{\frac{9+3\sqrt{9-80\mu^2}}{40\mu^2}}$ & 0 & 0 & 1 & $-\frac{3}{4\sqrt{5}}\leq\mu\leq\frac{3}{4\sqrt{5}}, \mu\neq0$ & $-1$ \\
\hline\hline
\end{tabular}
\end{adjustbox}
\caption{The fixed points and cosmological parameters
of the system Eqs.~(\ref{eqx})--(\ref{eqomega}) with $\Delta=1$ and $\Gamma=1$.}
\label{table:nonlin1} 
\end{table}
\begin{table}[t!]
\centering 
\vspace{3mm}
\begin{adjustbox}{width=\textwidth}
\begin{tabular}{c| c | c} 
\hline\hline  
Points & Eigenvalues  &  Stability\\ [0.5ex] 
\hline 
$A_1^\pm$ & $\left\{ -\frac{3}{2}, -1,  -\frac{3}{4} \left(1+\sqrt{\frac{4}{\lambda^2}-7}\right),  -\frac{3}{4}\left(1- \sqrt{\frac{4}{\lambda^2}-7}\right) \right\}$ & stable for: $1/2 < \lambda^2 \leq 4/7$,  \\ 
& & stable-focus for: $\lambda^2 > 4/7$, saddle for: $\lambda^2<1/2$ \\ 
$A_2^\pm$ & $\left\{-2, +1, -\frac{1}{2}\left(1+\sqrt{\frac{32}{3\lambda^2}-15}\, \right), -\frac{1}{2}\left(1-\sqrt{\frac{32}{3\lambda^2}-15}\, \right) \right\}$   & saddle. \\
$A_3^\pm  $ & $\left\{ -3, +3, +2, 3\pm3\lambda \right\}$  & saddle. \\  
$A_4$ & $\left\{ -\frac{3}{2}, -\frac{3}{2}, +\frac{3}{2}, -1 \right\}$ & saddle. \\ 
$A_5$ & $\left\{ -2, +2, -1, +1 \right\}$ & saddle. \\ 
$A_6^\pm$ & $\left\{-3 \lambda ^2,3 \left(\lambda ^2-1\right),6 \lambda ^2-3,6 \lambda ^2-4\right\}$ & stable for: $0<\lambda^2<1/2$; otherwise saddle. \\ 
\hline
$A_7^\pm$ & $\left\{ -4, -3, -\frac{3}{2}\left(1-\sqrt{1+\frac{32\lambda^2\mu^2-48\lambda^3\mu}{4\mu^2-\lambda^2(9-36\mu^2)}} \right), -\frac{3}{2}\left(1+\sqrt{1+\frac{32\lambda^2\mu^2-48\lambda^3\mu}{4\mu^2-\lambda^2(9-36\mu^2)}} \right) \right\}$ & stable for: $\text{con}_1$,  \\
$A_8^\pm$ & $\left\{ -4, -3, -\frac{3}{2}\left(1-\sqrt{1+\frac{32\lambda^2\mu^2-48\lambda^3\mu}{4\mu^2-\lambda^2(9-36\mu^2)}} \right), -\frac{3}{2}\left(1+\sqrt{1+\frac{32\lambda^2\mu^2-48\lambda^3\mu}{4\mu^2-\lambda^2(9-36\mu^2)}} \right) \right\}$ & stable focus for: $\frac{3\lambda^2}{-8\lambda^3+2\sqrt{16\lambda^6 + 17\lambda^4+\lambda^2}}<\mu< \frac{3\lambda}{2\sqrt{1+9\lambda^2}}$\,,\\
& & saddle for: 
\\
& & $\mu>\frac{3}{2}\sqrt{\frac{\lambda^2}{1+9\lambda^2}}$ if $\lambda<0$ or $\mu<-\frac{3}{2}\sqrt{\frac{\lambda^2}{1+9\lambda^2}}$ if $\lambda>0$. \\
$A_9^\pm$ & $\left\{ -\frac{3\lambda}{20\mu}\left(3-\sqrt{9-80\mu^2}\right),-4,-3,-3\right\}$ & stable for:  $\text{con}_2$; otherwise saddle.  \\ 
$A_{10}^\pm$ & $\left\{-4,-3,-3, -\frac{3\lambda}{20\mu}\left(3+\sqrt{9-80\mu^2}\right)\right\}$ & stable for:  $\text{con}_2$; otherwise saddle.  \\ [1ex] 
\hline\hline
\end{tabular}
\end{adjustbox}
\caption{The eigenvalues and the stability analyses of fixed points  the system Eqs.~(\ref{eqx})--(\ref{eqomega})}.
\label{table:nonlin2}
\end{table} 

Table~\ref{table:nonlin2} shows that the system is stable at the $A^{\pm}_{7,8}$ fixed points if the following conditions are satisfied: 
\begin{align} \label{eq:con1}
    \text{con}_1 &= \left\{\begin{array}{@{}lr@{}}
    0 < \mu \leq \frac{3\lambda^2}{-8\lambda^3+2\sqrt{16\lambda^6 + 17\lambda^4+\lambda^2}}, & \text{for } \lambda < 0\,\\ 
    -\frac{3\lambda^2}{8\lambda^3+2\sqrt{16\lambda^6+17\lambda^4+\lambda^2}} \leq \mu<0, & \text{for } \lambda > 0
    \end{array}\right.\,,
\end{align}
and also at the $A_{9}^\pm$ and $A_{10}^\pm$ fixed points if the following conditions are satisfied: 
\begin{align}
    \text{con}_2 &= \left\{\begin{array}{@{}lr@{}}
    -\frac{3}{4\sqrt{5}} \leq \mu < 0\,, & \text{for } \lambda < 0\\
    0 < \mu \leq \frac{3}{4\sqrt{5}}\,, & \text{for } \lambda >0
    \end{array}\right.\,. \label{eq:con2}
\end{align}
For obtaining Eq.~(\ref{eq:con1}), we take the existence conditions of the $A^{\pm}_{7,8}$ points into account, which implies the $\mu$ and $\lambda$ parameters should have the opposite signs, see the existence column of $A^\pm_{7,8}$ in Table~\ref{table:nonlin1}. The stability conditions in Eqs.~(\ref{eq:con1}) and (\ref{eq:con2}) put constraints on the $\{\lambda,\mu \}$ values. \emph{For example}, from Eq.~(\ref{eq:con1}), one can calculate that the largest (smallest) value for $\mu$ is $\mu_\text{max}\simeq 0.2271$ ($\mu_\text{min}\simeq -0.2271$) when $\lambda \simeq -0.3357$ ($\lambda \simeq 0.3357$). In both the small and the large $|\lambda|$ limit, the range of $\mu$ shrinks and eventually approaches to zero from both the negative and the positive sides. On the other hand, from Eq.~(\ref{eq:con2}), we see that the range of $\mu$ is determined depending on the sign of $\lambda$, but not on the amplitude. 
By taking these constraints on the numerical values of $\{\lambda, \mu\}$ into account, we plot the phase portraits of our dynamical system in Figure~\ref{fig:fig1}.
The blue points in the figure indicate the GB fixed points; namely the $A_{7,8}^\pm$, $A_{9}^\pm$, and  $A_{10}^\pm$ fixed points, and the numerical values of $\{\lambda, \mu\}$ are given such a way that most trajectories, including the red and green ones, are attracted to these stable points.  The gray shaded regions represent the region where the EoS parameter $\omega_\text{eff}=\omega_\phi<-1/3$ such that the accelerated expansion of the universe is possible. The dashed lines show the divergence lines given in Eq.~(\ref{eq:divergence}). 

\begin{figure}[t!]
    \centering
    \includegraphics[width=1\textwidth]{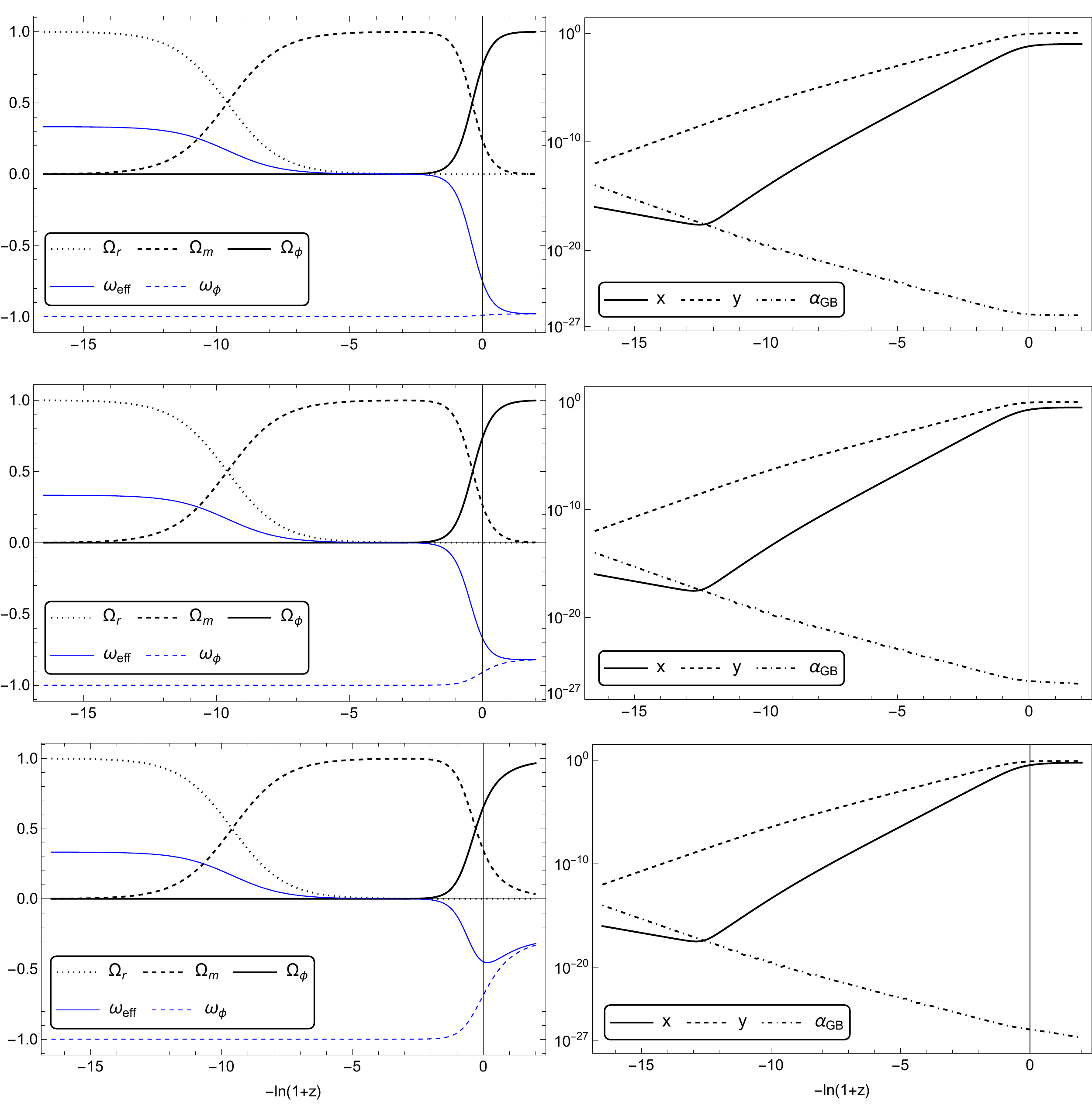}
    \caption{Left: The red-shift evolution of the $\Omega_i(z)$, where $i=r, m, \phi$, the EoS parameter $\omega_\phi$, and the effective EoS of the system $\omega_\text{eff}$. Right: The red-shift evolution of the $x, y$, $\alpha_\text{GB}$. In both columns, the $\lambda$ value increases from top to bottom; $\lambda=0.1, 0.3,$ and $0.6$, respectively. }
    \label{fig:fig2}
\end{figure}

The right column of Figure~\ref{fig:fig2} shows the time evolution of $\Omega_{i}(z) (i=r, m, \phi)$ and the EoS parameter of the scalar field $\omega_\phi(z)$, as well as the effective EoS of the system $\omega_\text{eff}(z)$, while  the left one shows the time evolution of the autonomous variables, $\{x(z),y(z),\alpha_\text{GB}(z)\}$. In the figure, from top to down, the numerical value of $\lambda$ changes as $\lambda=0.1, 0.3$, and $0.6$ for the $\mu=0.3$. If we change the $\mu$ value within the range given in Eqs.~(\ref{eq:con1}) and (\ref{eq:con2}) for each given $\lambda$, there is no noticeable change occurs because the $\mu$ value is too small to make significant contributions. In other words, we find that the cosmic phase transitions, as well as the time evolution of the autonomous variables, are more sensitive to the $\lambda$ value than the $\mu$ value as long as the stability conditions of the GB fixed points are concerned. 

The initial conditions of the autonomous variables are given in such a way that the universe experiences the right $\Omega_r \rightarrow \Omega_m \rightarrow \Omega_{\phi}$ phase transitions. Thus, the effective EoS parameter starts evolving from $\omega_{\text{eff}} = 1/3$ to a some value $\omega_{\text{eff}} < -1/3$, for allowing the late-time universe to experience the accelerated expansion. To replicate the viable cosmic history, the time evolution of $\{x,y,\alpha_{\text{GB}}\}$ must depend on fine-tuned initial conditions: $y > \alpha_{\text{GB}} > x$ in our case.
We show that at the onset of the radiation-dominated phase, the contribution of the GB term must be the second largest after the potential energy contribution of the scalar field, and its time evolution $\alpha_{\text{GB}}(z)$ experiences a continuous decrease over time. As a result, the present-day value of $\alpha_\text{GB}(z=0)$ is negligible, and so are the effects of the GB term. On contrary, the initial value of $x$, or the kinetic energy contribution of the scalar field, starts off as the smallest in the radiation-dominated phase, but its time evolution experiences a drastic increment during the matter-dominated phase after a short decrease towards the end of radiation dominated era, and it eventually catches up with the potential energy contributions, the $y(z)$, in the present universe. Thus, from Figure~\ref{fig:fig2}, it is explicit that the potential energy contribution of the scalar field plays the most dominant role throughout the cosmic history of the universe, and its present-day value is also the largest among the others. For the given $\mu$, Figure~\ref{fig:fig2} also shows that the present-day value of the EoS parameter increases as the $\lambda$ value increases, and it eventually surpasses ``$-1/3$'', ending the scalar field-dominated phase. In our case, the universe experiences the late-time acceleration, driven by the scalar field coupled to the GB term, if $\lambda<0.746$. 
\begin{figure}[t!]
    \centering
    \includegraphics[width=0.6\textwidth]{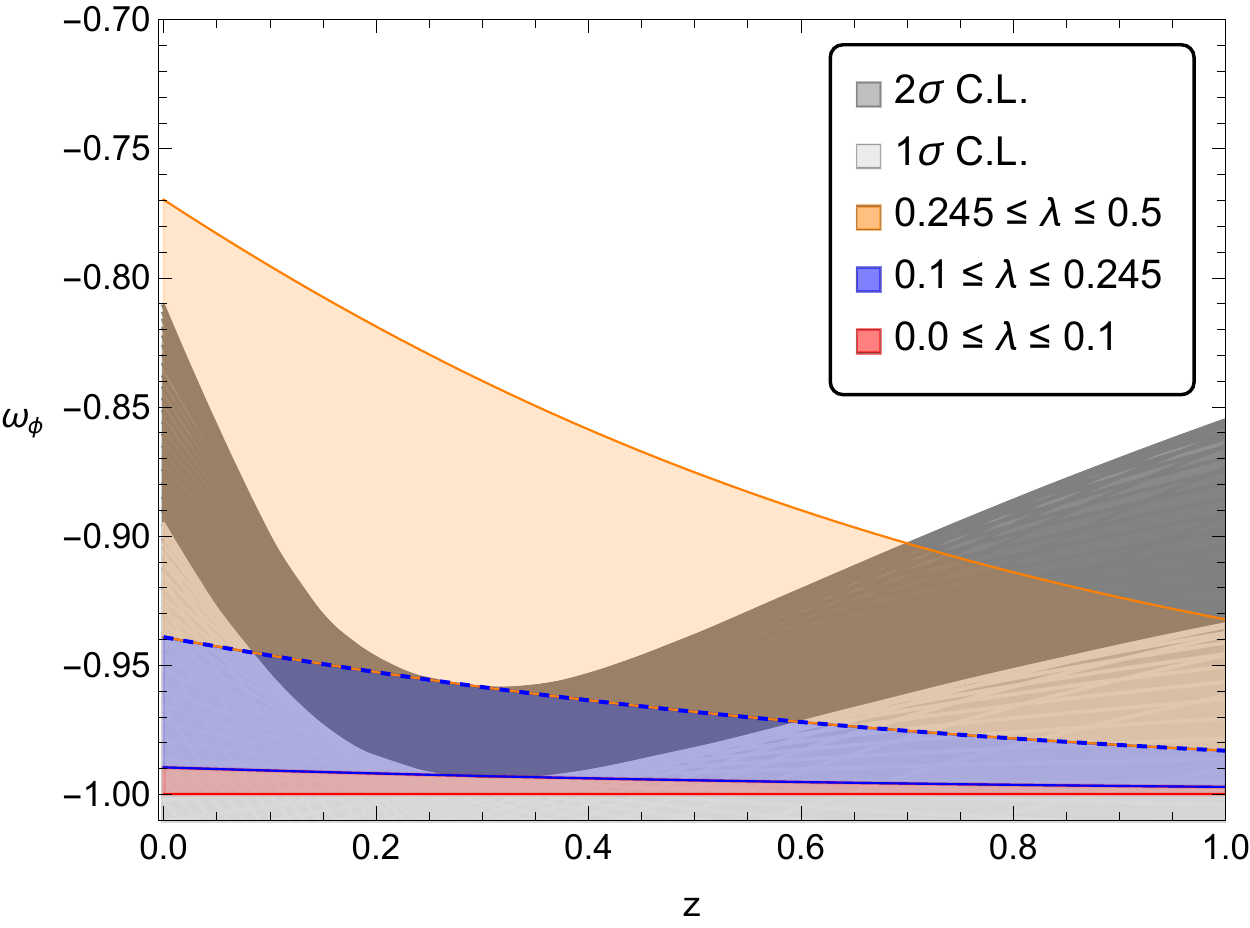}
    \caption{The red-shift evolution of the $\omega_\phi$ for different values of $\lambda$. The background dark- and light-gray shaded regions represent the observational bounds from the CMB, BAO, SnIa, and $H_0$ data using the CPL parameterization~\cite{Pan-STARRS1:2017jku}. The colored regions show the different parameter spaces of $\lambda$.  }
    \label{fig:fig3}
\end{figure}
To put further constraint on the potential parameter $\lambda$ in light of the observational bounds from CMB, BAO, SnIa, and $H_0$ data \cite{Pan-STARRS1:2017jku}, we plot in the red-shift evolution of the EoS $\omega_{\phi}(z)$ in Figure~\ref{fig:fig3}. In the figure, the parameter space for $\lambda$ is represented by the different colors and we adopted the Chevallier-Polarski-Linders (CPL) parameterization of the EoS that reads~\cite{Linder:2002et}, 
\begin{equation}\label{eq:CPLDarkEnergyEoS}
    \omega(z) = \omega_{0} + \frac{z}{1+z}\omega_a.
\end{equation} 
The result confirms our observation of Figure~\ref{fig:fig2} that the smaller values of $\lambda$ are more preferred. For instance, the preferred parameter space for $\lambda$ is $\lambda\lesssim0.3166$ to be consistent with the observational data at the $1\sigma$ level at present. The range is tighter being $\lambda\lesssim0.1$ to be consistent with the data for red-shift up to $z=1$.   
Thus, now we can provide tighter constraints on the parameters of our model given in Eq.~(\ref{eq:potandcfun}). In order for the universe to be stable at the GB fixed points and experience the late-time accelerated expansion, the parameters $\{\lambda,\mu\}$ of our model must be within the following range:
\begin{align}
    &\text{For } A^\pm_{7,8}: \quad -0.2268\lesssim \mu<0 \quad \text{when} \quad 0 < \lambda\leq0.3166\,,\label{eq:A78}\\
    &\text{For } A^\pm_{9,10}: \quad 0<\mu\lesssim\frac{3}{4\sqrt{5}} \quad \text{when} \quad 0 < \lambda\leq0.3166\,.\label{eq:A910}
\end{align}

\newpage
\section{Conclusions}\label{sec:sec5}
In this work, we have extended regularized $4D$ EGB gravity by introducing the non-minimal coupling function and the regularization scheme introduced in \cite{Glavan:2019inb}.  Along this consideration, a redefinition $\xi(\phi) \to \xi^{(D-4)}(\phi)$ is required so that EoM is well-defined. In the spatially flat space-time dimensions of FRW metric, we can write the EoM in arbitrary $D$-dimensions and then take the $D\to4$ limit.  As a solution, we derived the extended version of EoM obtained in \cite{Glavan:2019inb} for non-trivial $\xi(\phi)$, which is often associated with the cubic term in Hubble parameter and could be interesting in a further study of the cosmic evolution and the observational data constraint.

To explore the cosmological dynamics from this extension, we applied the dynamical system approach and analyzed the stability of critical fixed points for given exponential forms of the potential and the coupling function of the scalar field in Eq.~(\ref{eq:potandcfun}). We obtained ten fixed points, of which $A^\pm_1$ to $A^\pm_{6}$ are called the GR fixed points and $A^\pm_7$ to $A^\pm_{10}$ are called the GB fixed points because their existence is due to the presence of non-minimal coupling function $\xi(\phi)\rightarrow\xi^{(D-4)}(\phi)$ and rescaling of the coupling constant $\alpha\to \alpha/(D-4)$, and they exist and stable under specific conditions listed in Table~\ref{table:nonlin1} and~\ref{table:nonlin2}. The GB fixed points correspond to the de-Sitter solution with the EoS parameter $\omega_{\text{eff}}=-1=\omega_\phi$ that admits the late-time accelerating universe under the conditions in Eqs.~(\ref{eq:con1}) and (\ref{eq:con2}). The phase space plots in Figure~\ref{fig:fig1} confirm the stable and accelerating solutions.

We also present the cosmic history of the universe, $\Omega_r \rightarrow \Omega_m \rightarrow \Omega_{\phi}$, in Figure~\ref{fig:fig2} along with the time evolution of dynamical variables; $x$, $y$, and $\alpha_\text{GB}$, associated with the kinetic and potential energy contribution of the scalar field and the GB contribution, respectively. Both kinetic and potential energies of the scalar field play the most important role in the present day; particularly the potential energy, while the effect of the GB term is appreciated at the beginning of the radiation-dominated epoch. However, the GB contribution decreases in time and, eventually, it does not play many roles in the present universe. 

The time evolution of the EoS parameter for different values of $\lambda$, the potential parameter, is presented in Figure \ref{fig:fig3} together with the upper bounds from the observational data. The effective EoS starts evolving from $\omega_{\text{eff}}=1/3$ to some value $\omega_{\text{eff}}<-1/3$, which indicates the late-time universe experiences the accelerated expansion. If $\lambda<0.746$, our model explains the late-time accelerated expansion of the universe driven by the scalar field coupled to the GB term  and the EoS satisfies the observational constraint by CMB, BAO, SnIa, and $H_0$ data if $\lambda \lesssim 0.1$ at $1\sigma$ C.L.. In conclusion, our model suggests that the universe can be stable at the GB fixed points and experiences the late-time accelerated expansion when the values of the model parameters $\{\lambda,\mu\}$ are in the range given in Eqs. (\ref{eq:A78}) and (\ref{eq:A910}).    

\section*{Acknowledgments}
BB is partially supported by the project "Extension of the Standard Model and Cosmology, heavy meson study" (SHUTBIHHZG-2022/168) by the Mongolian Ministry of Education and Science and Mongolian Science and Technology Foundation. SK is supported by the Higher Education Improvement Project (HEIP) funded by the Cambodian Government (IDA Credit No.~6221-KH) and PR is supported by the Swedish International Development Cooperation Agency (SIDA) through Sweden and the Royal University of Phnom Penh (RUPP)'s Pilot Research Cooperation Programme (Sida Contribution No.~11599). GT was supported by Basic Science Research Program through the National Research Foundation of Korea(NRF) funded by the Ministry of Education (grant number) (NRF-2022R1I1A1A01053784) and (NRF-2021R1A2C1005748).



\end{document}